\documentclass[11pt,a4paper,aps,tightenlines,
preprintnumbers,nofootinbib,showkeys,superscriptaddress]{revtex4-1}
\pdfoutput=1
\usepackage{fullpage}
\usepackage{float}
\usepackage{amsfonts}
\usepackage{amsmath}
\usepackage{slashed}
\usepackage{mathrsfs}
\usepackage[scr=rsfs,cal=boondox]{mathalfa}
\usepackage{amssymb}
\usepackage{graphicx}
\usepackage{makeidx}
\usepackage{cancel}
\usepackage{eepic}
\usepackage{epsfig}
\usepackage{latexsym}
\usepackage{mathtools}
\usepackage[dvipsnames]{xcolor}
\usepackage{float}
\usepackage{multirow}
\usepackage[export]{adjustbox}
\usepackage{xurl,hyperref}
\usepackage{enumitem}
\hypersetup{colorlinks=true,citecolor=red,linkcolor=NavyBlue,urlcolor=NavyBlue}
\usepackage[utf8]{inputenc}
\usepackage[caption=false]{subfig}
 
\usepackage{natbib}
\usepackage{mathptmx}

\def\ba{\begin{array}}       
\def\ea{\end{array}}

\def\beq{\begin{eqnarray}}
\def\eeq{\end{eqnarray}}
\begin{document}
\title{Production of Singlet dominated scalar(s) at the LHC}
\author{Subhadip Bisal}
\email{subhadip.b@iopb.res.in}
\affiliation{Institute of Physics, Sachivalaya Marg, Bhubaneswar 751 005, India}
\affiliation{Homi Bhabha National Institute, Training School Complex, Anushakti Nagar, Mumbai 400 085, India}
\author{Debottam Das}
\email{debottam@iopb.res.in}
\affiliation{Institute of Physics, Sachivalaya Marg, Bhubaneswar 751 005, India}
\affiliation{Homi Bhabha National Institute, Training School Complex, Anushakti Nagar, Mumbai 400 085, India}
\author{Swapan Majhi}
\email{majhi.majhi@gmail.com}
\affiliation{Ranaghat College, Nadia, West Bengal-741201, India.}

\author{Subhadip Mitra}
\email{subhadip.mitra@iiit.ac.in}
\affiliation{Center for Computational Natural Sciences and Bioinformatics, International Institute of Information Technology, Hyderabad 500 032, India}

\date{\today}
\preprint{IP/BBSR/2022-04}

\begin{abstract}
\noindent
 The leading order production of an SM singlet-like scalar has primarily been realized through the gluon fusion process by mixing with the $SU(2)_L$ scalar doublet of the model. The dominant part of the physical state, i.e., the singlet component, does not have any role in its direct production. Focusing on such a state with a mass smaller than the SM-like Higgs scalar, we calculate the dominant next-to-leading (NLO) order corrections to its production cross-section. With these improved cross-sections, the present and future LHC limits may become somewhat more stringent.
\end{abstract}

\maketitle
\section{INTRODUCTION}
The Higgs boson was discovered a few years ago
\cite{Aad:2012tfa,Chatrchyan:2012xdj,Aad:2019mbh} at the Large Hadron Collider (LHC) with Standard-Model (SM)-like interactions and mass $m_h \simeq 125~$GeV. 
In various beyond the SM (BSM) scenarios, it often appears as a part of an enlarged scalar sector that also contains additional Higgs states. There are various possibilities with additional scalars that are singlet under the SM gauge group but may be involved in the electroweak symmetry breaking (EWSB) through renormalizable interactions with the Higgs doublets. 
In this paper, we consider the minimal possible extension where a gauge-singlet scalar is added  
to the SM particle content~\cite{Silveira:1985rk,McDonald:1993ex,Burgess:2000yq,McDonald:2001vt,Schabinger:2005ei,O'Connell:2006wi,Bowen:2007ia,Barger:2007im,He:2008qm,Cline:2013gha,Das:2020ozo}. A singlet-like state is often considered to explain the dark matter abundance~\cite{McDonald:2001vt}, 
initiate a first-order electroweak phase transition (EWPT)~\cite{Profumo:2007wc}, or generate the neutrino masses \cite{Mohapatra:1986bd,De:2021crr}. In the context of Supersymmetry (SUSY),
the Next to Minimal Supersymmetric Standard Model (NMSSM) contains a singlet superfield \cite{Ellwanger:2009dp, Maniatis:2009re}. The singlet superfield is helpful in addressing the $\mu$ problem in the Minimal Supersymmetric Standard Model (MSSM).
\vskip 0.15cm

Since a singlet-like scalar is well motivated in a wide range of BSM scenarios, it is important to study its production and possible signatures at the LHC. At the LHC, a spin-$0$ state can
either be produced directly through the gluon fusion process (ggF) at one-loop or via the cascade decays of some heavier states. Its cross-section may be computed in terms of its doublet component. For example, when the SM is augmented with a real singlet scalar $\varphi$ (from here on, we refer to this model as the $\varphi$+SM scenario), one obtains an SM-like $h$ and a mostly singlet state $\phi$ in the physical basis. The leading order (LO) cross-section of $\phi$ essentially goes as the production cross-section of the doublet scalar at mass $m_{\phi}$ but is suppressed by the tiny doublet component inside. As a result, the LO $\phi$ production cross-section for similar masses is smaller than that of $h$. Similarly, the tiny doublet part generally determines the cascade productions of $\phi$ as well.
As the LHC bounds on the couplings of $h$ to the gauge bosons and fermions are expected to become tighter~\cite{Cepeda:2019klc}, the mixing between the singlet and the doublet scalars will be severely restricted. Consequently, producing the singlet $\phi$ at the LHC would become more and more challenging.
We note that the new state may also open up a possibility where the vector boson fusion may become important \cite{Das:2018fog} in the production of a spin-0 state.

\vskip 0.15cm

In this letter, going beyond the LO calculation, we consider the next-to-leading order (NLO) electroweak (EW) correction to $\phi$ production. It is induced by a tree-level $\phi h h$ vertex from the renormalizable Higgs-portal-like $\varphi-H$ interactions. This results in a non-vanishing $ g g \phi$ coupling at the two-loop level (see Fig.~\ref{fig:ggphi_two_loop}). Earlier, the two-loop electroweak processes mediated by fermions and gauge bosons were computed in \cite{Aglietti:2004nj,Degrassi:2004mx,Degrassi:2016wml} (also see \cite{Dawson:2013uqa}) which lead to an overall $\delta_{EW}\sim$ $5\%$ contribution at the NLO to an SM-like Higgs scalar $h$. Here, we consider the corrections for a dominantly
singlet-like state, where the contributions mediated by the EW gauge bosons would
become less important. 
As we will see, overall, an enhancement up to $\sim 7\%$  to the LO cross-section may be observed. Here, our primary concern is a
light $\phi$ ($m_\phi \ll m_h$), and our analysis can be generalized to any model with an extended scalar sector.

\section{The gluon fusion cross-section of a Singlet-like state }
\label{sec:frame}
\begin{figure}[!ht]
		\centering
\includegraphics[width=0.85\linewidth]{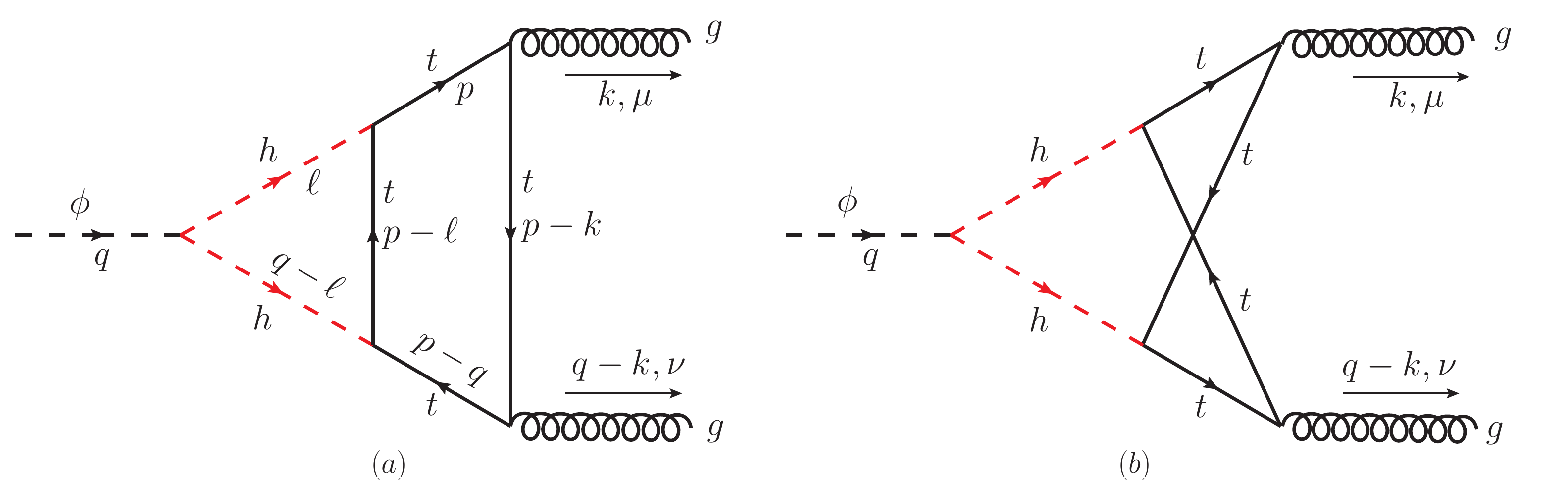}
\caption{Two-loop contributions to the $g g \phi $ coupling, mediated by the $\phi hh$ vertex.}
\label{fig:ggphi_two_loop}
\end{figure}
We now look at the amplitude for the relevant two-loop EW processes, 
$g^{A,\mu} (k) + g^{B,\nu} (q-k) \to \phi$,  mediated by the $\phi hh$ cubic interaction, as depicted in Fig.~\ref{fig:ggphi_two_loop}. We calculate the above two-loop diagrams with the help of one-loop effective vertices.\footnote{A similar technique was earlier used by Barr and Zee in calculating the dipole moments of the electron and the neutron \cite{Barr:1990vd}.} Such a disentanglement can be conceived from the general two-loop integral in terms of two momentum variables. We  discuss it in detail in the next sections.

\subsection{CALCULATION OF THE TWO-LOOP  $gg\phi$ AMPLITUDE}
\label{sec:decompose}
\begin{figure}[H]
	\centering
	\includegraphics[width=1.0\linewidth]{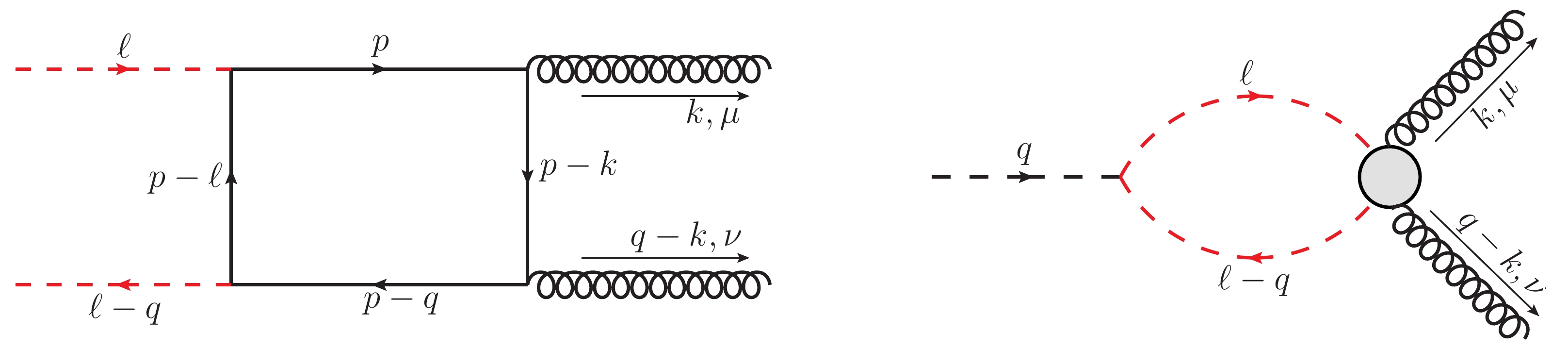}
	\qquad\qquad(a)\qquad\qquad\qquad\qquad\qquad\qquad\qquad\qquad\qquad\qquad\qquad(b)\qquad
	\caption{Realization of the diagram \ref{fig:ggphi_two_loop}a ({\it planar} diagram) through one-loop effective vertices.}
	\label{fig:1stdiagram}
\end{figure}

To appreciate the method that
would be followed in the analysis, we start
with the two-loop {\it planar} vertex (Fig.\ref{fig:ggphi_two_loop}a). Its
amplitude, $\mathcal{M}^{\mu\nu}$, can be expressed as,
\begin{align}
-i\mathcal{M}^{\mu\nu} = Q \,{\rm Tr}\int \frac{d^dp}{(2\pi)^d}\int  \frac{d^d\ell}{(2\pi)^d}\Bigg[\frac{\gamma^\mu (\slashed{p}+m_t)(\slashed{p}-\slashed{\ell}+m_t)(\slashed{p}-\slashed{q}+m_t)\gamma^\nu (\slashed{p}-\slashed{k}+m_t)}{D_{1...6} }\Bigg]~,
\label{fulltwoloop_1}
\end{align}
where $D_i (i=1...6)$ 
read as $D_1 = \ell^2-m_h^2$, $D_2=(q-\ell)^2-m_h^2$, 
$D_3 = (p-\ell)^2-m_t^2$, $D_4 = p^2-m_t^2$, $D_5 = (p-q)^2-m_t^2$, and $D_6 = (p-k)^2-m_t^2$. 
We calculate the prefactor $Q$  explicitly below. 
Note that, only $D_1, D_2,$ and $D_3$ are the $\ell$-dependent denominators. 
The loop integral in Eq.~\eqref{fulltwoloop_1} can be 
expressed in a more intuitive form,
\begin{align}
-i \mathcal{M}^{\mu\nu}= \Lambda\, \int \frac{d^d\ell}{(2\pi)^d}\Bigg[\frac{\Xi_{1}^{\mu\nu}}{D_1D_2}\Bigg]~,
 \label{eq:phiggg1}
\end{align}
where $\Lambda$ is the $\phi hh $ coupling, and 
\begin{align}
 \Xi_{1}^{\mu\nu} = \mathcal{S} ~{\rm Tr}\int \frac{d^dp}{(2\pi)^d} \Bigg[\frac{\gamma^\mu (\slashed{p}+m_t)(\slashed{p}-\slashed{\ell}+m_t)(\slashed{p}-\slashed{q}+m_t)\gamma^\nu (\slashed{p}-\slashed{k}+m_t)}{D_{3...6} }\Bigg]~.
 \label{eq:gghh1}
\end{align}
Following Eq.~\eqref{fulltwoloop_1},
$Q=\Lambda \mathcal{S}$, where $\mathcal{S}$ is 
calculated below. Notably, 
the amplitude in  Eq.~\eqref{eq:gghh1} can be expressed in terms of 
the one-loop $gghh$ effective vertex shown in Fig.~\ref{fig:1stdiagram}a. 
Subsequently, we  use $\Xi_{1}^{\mu\nu}$ in Eq.~\eqref{eq:phiggg1} for evaluating the $gg \phi$ amplitude shown in Fig.~\ref{fig:1stdiagram}b. 
This way, the two-loop amplitude can essentially be realized through the one-loop vertices $gghh$ and $gg \phi$, respectively.  
While evaluating the first loop, we keep the full momentum dependence for the two Higgs scalars and then execute the loop integral over the Higgs momentum to calculate the amplitude of $gg \phi$. 

\noindent
Now, the amplitude for the $gghh$ vertex in Fig.~\ref{fig:1stdiagram}a can be written as, 
\begin{align}
\Xi_1^{\mu\nu} = \mathcal{S} \int \frac{d^dp}{(2\pi)^d} \Bigg[\frac{\mathcal{N}_1^{\mu\nu}}{\mathcal{D}}\Bigg]~,
\end{align}
\noindent
where $\mathcal{S}=-{{i}}g_s^2{m_t^2}{\rm Tr}(T^aT^b)/v^2$ with $g_s$ being the strong coupling constant, $m_t$ the top-quark mass, and $T^{a,b}$ the generators of $SU(3)_c$. The numerator, $\mathcal{N}_1^{\mu\nu}$, can be read from Eq. \eqref{eq:gghh1} as, 
$\mathcal{N}_1^{\mu\nu} = {\rm Tr}\Big[\gamma^\mu (\slashed{p}+m_t)(\slashed{p}-\slashed{\ell}+m_t)(\slashed{p}-\slashed{q}+m_t)\gamma^\nu (\slashed{p}-\slashed{k}+m_t)\Big]$, and the denominator,  $\mathcal{D}=\Big[(p^2-m_t^2)\bigl\{ (p-\ell)^2-m_t^2\bigr\} \bigl\{ (p-q)^2-m_t^2\bigr\} \bigl\{ (p-k)^2-m_t^2\bigr\} \Big]$.

 Since gluons  are transverse in nature,  we use the transverse projection operator, $\mathcal{P}_{T,\mu\nu} = g_{\mu\nu}-\frac{k_\nu(q-k)_\mu}{k\cdot(q-k)}$ to write $\Xi_1^{\mu\nu}$ as 
 $\Xi_1^{\mu\nu}\sim \Xi_1 \mathcal{P}_{T,\mu\nu}$, where  $\Xi_1$ is the scalar $gghh$ vertex factor. $\Xi_1$ can be calculated by the following relation,
 \begin{align}
 \Xi_1 = \frac{1}{2} \Xi_1^{\mu\nu}\mathcal{P}_{T,\mu\nu}~,
 \label{scalar_vertex_factor}
 \end{align}  
where we have used the properties of $\mathcal{P}_{T,\mu\nu}$: $\mathcal{P}_{T,\mu\nu}k_1^\mu = 0 = \mathcal{P}_{T,\mu\nu}k_2^\nu$ and $\mathcal{P}_{T,\mu\nu}\mathcal{P}_T^{\mu\nu} = 2$ ($k_1^\mu$ and $k_2^\nu$ denote the gluon momenta). Therefore, one can recast Eq.~\eqref{scalar_vertex_factor} as,
\begin{align}
\Xi_1 = \mathcal{S} \int \frac{d^dp}{(2\pi)^d}\Big[\frac{\mathcal{N}_1^{\mu\nu}\mathcal{P}_{T,\mu\nu}}{\mathcal{D}}\Big]~.
\label{scalar_vertex_factor_1}
\end{align}
The prefactor $\mathcal{S}$ can be expressed in a conventional form, 
$\mathcal{S} = -i \pi \alpha_s {m_t^2}\delta^{ab}/{v^2}$, where we have 
used ${\rm Tr}(T^aT^b)=\delta^{ab}/2$ and $\alpha_s=g_s^2/(4\pi)$.
At this point, it is convenient
to simplify the denominator, assuming  
 vanishing momenta for the gluons and $\phi$, i.e., setting $k,q\rightarrow0$.
Then introducing the Feynman parametrization, Eq.~\eqref{scalar_vertex_factor_1} becomes
\begin{align}
\Xi_1 = \mathcal{S} \int_{0}^{1} dx \int \frac{d^d\mathcal{k}}{(2\pi)^d}\Bigg[\frac{3\mathcal{N}_1 (1-x)^2}{(\mathcal{k}^2-\Delta)^4}\Bigg]~,
\label{scalar_vertex_factor_1_2nd}
\end{align}
where $x$ is the Feynman parameter, $\mathcal{k}\equiv p-x\ell$ is the new shifted momenta, $\Delta=x(x-1)\ell^2 + m_t^2$ and $\mathcal{N}_1= \mathcal{N}_1^{\mu\nu}\mathcal{P}_{T,\mu\nu}$. 

\begin{figure}[H]
	\centering
	\includegraphics[width=1.0\linewidth]{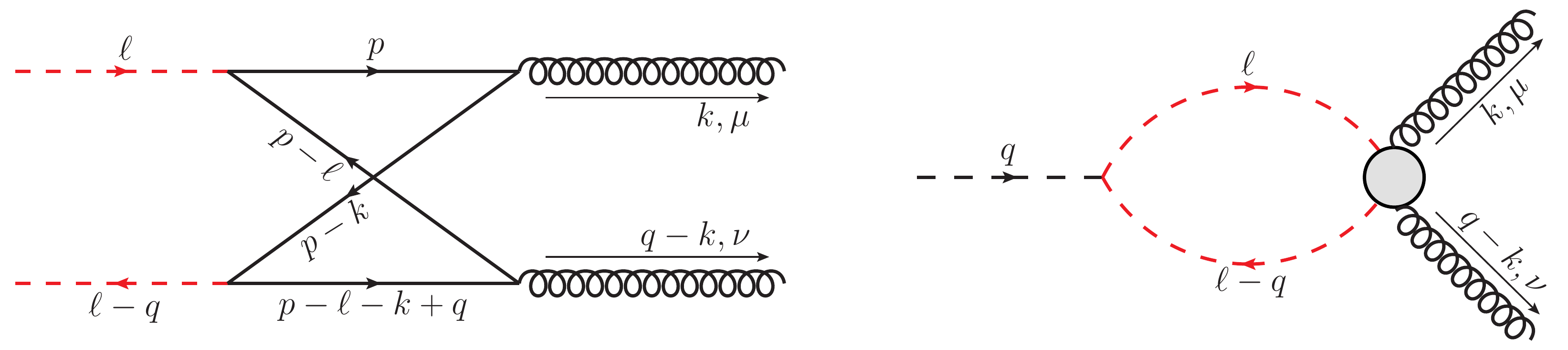}
	\qquad\qquad(a)\qquad\qquad\qquad\qquad\qquad\qquad\qquad\qquad\qquad\qquad\qquad(b)\qquad
	\caption{ Realization of the diagram \ref{fig:ggphi_two_loop}b ({\it non-planar} diagram) through one-loop effective vertices.}
	\label{fig:2nddiagram}
\end{figure}

In a similar manner,
the amplitude of the {\it non-planar} $gg \phi $ vertex (Fig.~\ref{fig:ggphi_two_loop}b) can also be realized through one-loop vertices, as shown in Fig.~\ref{fig:2nddiagram}a and Fig.~\ref{fig:2nddiagram}b. 
Then the scalar
form factor $\Xi_2$ for the $gghh$ effective vertex in Fig.~\ref{fig:2nddiagram}a reads as,
\begin{align}
\Xi_2 = \mathcal{S} \int_{0}^{1} dx \int \frac{d^d\mathcal{k}}{(2\pi)^d}\Bigg[\frac{6\mathcal{N}_2 x(1-x)}{(\mathcal{k}^2-\Delta)^4}\Bigg]~.
\label{scalar_vertex_factor_2}
\end{align}
The numerator in Eq.~\eqref{scalar_vertex_factor_1_2nd} and in Eq.~\eqref{scalar_vertex_factor_2} can be assembled  in a suitable basis:
\begin{align}
\mathcal{N}_i &= (\mathcal{k.k})\xi_1^{(i)} + (\mathcal{k.k})^2\xi_2^{(i)} + (k.\mathcal{k})(q.\mathcal{k})\xi_3^{(i)} + (k.\mathcal{k})^2\xi_4^{(i)} + (q.\mathcal{k})^2 \xi_5^{(i)} + (\mathcal{k.k})(k.\mathcal{k})(q.\mathcal{k})\xi_6^{(i)}\nonumber\\
& + (\mathcal{k.k})(k.\mathcal{k})^2 \xi_7^{(i)} + (\mathcal{k.k})(q.\mathcal{k})^2\xi_8^{(i)} + (q.\mathcal{k})(\ell.\mathcal{k})\xi_9^{(i)}+ (\ell.\mathcal{k})^2\xi_{10}^{(i)}+ (k.\mathcal{k})(\ell.\mathcal{k})\xi_{11}^{(i)} + \xi_{12}^{(i)}~.
\end{align}
Here, $\xi_1^{(i)},...,\xi_{12}^{(i)}$ ($i=1, 2$ stands for  
the diagrams in Fig.~\ref{fig:1stdiagram}a and \ref{fig:2nddiagram}a, respectively) are the scalar functions which depend on $m_t$, $m_\phi$, the Feynman parameter $x$, and scalar combinations of different momenta $k$, $q$, and $\ell$. They are defined 
in Appendix \eqref{appedxixA}. After performing the momentum integration over the shifted momentum variable $\mathcal{k}$, we can write the $gghh$ vertex in a compact form:
\begin{align}
\Xi_i = 3\mathcal{S} \int_{0}^{1}dx \,F^{(i)}\Bigg[\mathcal{H}_0^{(i)} + \Bigg(\frac{1}{\Delta}\Bigg)\mathcal{H}_1^{(i)} + \Bigg(\frac{1}{\Delta^2}\Bigg)\mathcal{H}_2^{(i)} \Bigg]~,
\label{effective_vertex_1}
\end{align}
where, $F^{(1)}=(1-x)^2$ and $F^{(2)}=2(1-x)x$. The different form factors are 
classified as follows.
\begin{align}
\mathcal{H}_0^{(i)} =& \frac{i}{(4\pi)^2}\Bigg[\Bigg(\Delta_\epsilon -\log\Delta-\frac{5}{6}\Bigg)\xi_2^{(i)} + \frac{m_\phi^2}{8}\Bigg(\Delta_\epsilon-\log\Delta-\frac{1}{3}\Bigg)\xi_6^{(i)}\Bigg]~,\\
\mathcal{H}_1^{(i)} =& -\frac{i}{(4\pi)^2}\Bigg[\frac{1}{3}\xi_1^{(i)} + \frac{m_\phi^2}{24}\xi_3^{(i)}+ \frac{m_\phi^2}{12}\xi_5^{(i)} + \frac{\ell.q}{12}\xi_9^{(i)}+ \frac{\ell.\ell}{12}\xi_{10}^{(i)}+ \frac{\ell.k}{12}\xi_{11}^{(i)}\Bigg]~,\\
\mathcal{H}_2^{(i)} =& \frac{i}{(4\pi)^2}\frac{1}{6} \xi_{12}^{(i)}~,
\end{align}
where $\Delta_\epsilon\equiv \frac{2}{\epsilon}-\gamma+ \log(4\pi)$, $d=4-\epsilon$, and $\gamma = \lim\limits_{n\rightarrow\infty}\Big(-\log n + \sum_{r=1}^{n}\frac{1}{r}\Big)\simeq0.57722$ is the Euler–Mascheroni constant.

As stated earlier, we can now use Eq.~\eqref{effective_vertex_1} to calculate the two-loop coupling between the singlet-like scalar $\phi$ and two gluons. In other words, we can substitute the effective $gghh$ amplitude 
in Fig.~\ref{fig:1stdiagram}b and \ref{fig:2nddiagram}b while keeping
full momentum dependence for the Higgs fields. The respective
amplitudes can be calculated as,
\begin{align}
\mathcal{M}_i = i\Lambda \int \frac{d^d\ell}{(2\pi)^d}\Bigg[\frac{\Xi_i}{(\ell^2-m_h^2)\left\lbrace(\ell-q)^2-m_h^2 \right\rbrace }\Bigg]~,
\label{phiggamplitude1}
\end{align} 
where $\Lambda$ is the $\phi hh$ coupling as before and $m_h$ is the mass of the SM-like Higgs scalar.
After performing the momentum integral over $\ell$, we write down the two-loop amplitude for the $gg \phi$ vertex:
\begin{align}
\mathcal{M}^{(\rm 2Loop)}_{ gg  \phi} = i\Lambda \mathcal{S}  \mathcal{M}
\label{eq:amplitude}
\end{align}
where $\mathcal{M}=\frac{i}{2^{12}\pi^4} (2\,\, \overline{\mathcal{M}}_1+ \overline{\mathcal{M}}_2)$ with $\overline{\mathcal{M}}_1$ and $\overline{\mathcal{M}}_2$ are the form factors arising from Eq.~\eqref{eq:amplitude}.
Note that there is a symmetry factor, $2$ for the diagram in Fig.~\ref{fig:ggphi_two_loop}a.\footnote{There is no symmetry factor for the {\it non-planar} diagram (see e.g., \cite{Degrassi:2016wml}).} Finally, the amplitudes in Eq.~\eqref{phiggamplitude1} can be evaluated using Package-X \cite{Patel:2015tea,Patel:2016fam}. The total amplitude $\mathcal{M}$ is completely UV-finite. Moreover, it is also independent of the choice of the renormalization scale. 
For the sake of completeness, we also show
the LO amplitude $\mathcal{M}^{\rm LO}$ for the production of an SM-like Higgs scalar, 
\begin{align}
    \mathcal{M}^{\rm LO} = \frac{1}{4\pi} (\sqrt{2}G_F)^{\frac{1}{2}} m_t^2\alpha_s \big[2 + (\tau-1 )f(\tau)\big]~,
    \label{EQ:LO}
\end{align}
where
\begin{align}
   f(\tau) =  
        -2 \arcsin^2\Bigg(\sqrt{\frac{1}{\tau}}\Bigg)~,\quad & \tau> 1~~;~~~
       \frac{1}{2}\Bigg[\log\Bigg(\frac{1+\sqrt{1-\tau}}{1-\sqrt{1-\tau}}\Bigg)-i\pi\Bigg]^2~, & \tau<1 
  \nonumber
\end{align}
and $\tau=\frac{4m_t^2}{m_h^2}$. For an SM-like Higgs, the ratio, $\frac {2 \mathcal M_{gg \phi }^{( \rm 2Loop)}} {\mathcal M^{\rm LO}}$ has been computed adopting a different route in Ref.~\cite{Degrassi:2016wml} earlier.
Now,
we 
also calculate the same ratio by replacing the coupling $\Lambda$ with the trilinear coupling for an SM-like Higgs scalar $\Lambda_{hhh}^{\rm SM}$ = $\frac{3 m_h^2}{v}$ with  $v= \rm 246 GeV$. We find that it closely agrees with the value obtained in Ref.~\cite{Degrassi:2016wml}.

\subsection{THE NLO CORRECTIONS TO CROSS-SECTION OF $\phi$}
\label{sec:calphicross}
We can calculate the NLO cross-section for the production of $\phi$ as follows.
\begin{align}
    \sigma_{\phi}^{\rm NLO} 
    =\sigma_{\phi}^{\rm LO} + \sigma_{gg\phi} + 2 \sqrt{\sigma_{gg\phi} \sigma_{\phi}^{\rm LO}}~,
    \label{Eq:nlo_lo_phi}
\end{align}
\noindent
where $\sigma_{\phi}^{\rm LO}$ is the LO cross-section including the QCD corrections in the SM. In addition to the LO value, $\sigma_{\phi}^{\rm NLO}$ includes contribution linear in $\Lambda$ from the interference between the LO amplitude $\mathcal M^{\rm LO}$ and $\mathcal M_{gg \phi }^{(\rm 2Loop)}$. 
Since the contribution from the wave-function renormalization, $\delta Z_{\phi}$ \cite{Degrassi:2016wml} 
is small for our choice of parameters (described in the next sections), we ignore it in Eq. \eqref{Eq:nlo_lo_phi}.

In the effective-theory description, the interaction generated by
$\mathcal{M}_{ gg \phi}^{(\rm 2Loop)}$  
can be expressed in terms of the gluon field-strength tensor, $G_{\mu\nu} = \partial_{\mu}g_{\nu}- \partial_{\nu}g_{\mu}+\mathbb{O}(g^2)$.
Switching from the position space to the momentum space ($\partial \rightarrow ik$), we can write
\begin{align}
G_{\mu\nu}G^{\mu\nu} \rightarrow i\left(
k_{\mu}g_{1\nu}-k_{\nu}g_{1\mu}\right) i\left( (q-k)^{\mu}g^{2\nu}-(q-k)^{\nu}g^{2\mu}\right) + \mathbb{O}(g^{3})=m_{\phi}^{2}g_{1\mu}g_{2\nu}P_{T}^{\mu\nu} + \mathbb{O}(g^{3}) ,
\label{position_momentum_space_trn}
\end{align}
where $g_{1\mu}$ and $g_{2\nu}$ are the field vectors for the two gluons respectively. 
We can express the $gg \phi$ interaction in terms of an effective Lagrangian:
\begin{align}
\mathcal{L}_{\rm eff.} = \frac{1}{v} g_{gg \phi } \phi G_{\mu\nu}G^{\mu\nu}~.
\label{ggphi_coupling}
\end{align}	
In the above, we can split the effective coupling $g_{ gg \phi}$ into two parts, $g_{ gg \phi} = g^D_{ gg \phi} + g^S_{gg \phi }$. The LO $\phi$ production goes through $g^D_{ gg \phi}$, which essentially connects with the tiny doublet component of $\phi$ at the one-loop level, and $g^S_{ gg \phi}$ leads to the two-loop $ g g\to\phi$ amplitude. We find that,
\begin{align}
g^S_{ gg \phi}= \frac{v \Lambda \mathcal{S} \mathcal{M}}{m_\phi^2}
\quad {\rm and }\quad g^D_{ gg \phi}= \mathcal{K}_D~g_{ggh}~,
\label{Eq:effcoup}
\end{align}
where $\mathcal{K}_D$ is the reduced coupling of $\phi$ and $g_{ggh}$ is the SM-like Higgs coupling to gluons. We have evaluated $g_{ggh}$ using Eqs.~\eqref{EQ:LO} and~\eqref{ggphi_coupling}.\footnote{To evaluate $g_{ggh}$, one has to replace Eq.~\eqref{ggphi_coupling} by the effective Lagrangian, $\mathcal{L}_{\rm eff.}=\dfrac{1}{v}g_{ggh}hG_{\mu\nu}G^{\mu\nu}$.}
Additionally, one can include the higher-order effects through a $K$-factor which we choose to be $1$ for simplicity. 

Since we are interested in the contributions for a dominantly singlet-like state, it is instructive to note the partial decay width of $\phi$ from the process $\phi \rightarrow gg$, mediated by $g^S_{ g g \phi}$ in particular,
\begin{align}
\Gamma( \phi \rightarrow gg) = \Bigg(\frac{\Lambda \mathcal{S} \mathcal{M}}{m_\phi^2}\Bigg)^2 \frac{2m_\phi^3}{\pi}~.
\label{phi_decay_width}
\end{align} 
The partonic cross-section of the $gg\to\phi$ process 
reads as,
\begin{align}
    \hat{\sigma}(gg\rightarrow \phi) = \frac{\pi^2 m_\phi}{8\hat{s}}\Gamma_{ \phi \rightarrow gg} \delta(\hat{s}-m_{\phi}^2).
    \label{ggphi_cross_sec}
\end{align}
One can 
calculate the hadronic cross-section by using Eq.~\eqref{ggphi_cross_sec} and the gluon parton distribution function (PDF) in the following way:
 \begin{align}
    \sigma_{gg\phi} &= \int_{0}^{1}dz_1 \int_{0}^{1}dz_2 \,\mathcal{G}(z_1) \mathcal{G}(z_2) \hat{\sigma}(gg\rightarrow\phi)\nonumber\\ 
    &= \sigma_0 m_\phi^2 \int_{0}^{1}dz_1 \int_{0}^{1}dz_2 \,\mathcal{G}(z_1) \mathcal{G}(z_2)\delta(z_1 z_2 \mathbf{S} - m_\phi^2)\nonumber\\
    &= \sigma_0 \tau \frac{d\mathcal{L}^{gg}}{d\tau}~,
    \label{eq:had}
 \end{align}
where $\sigma_0 = \frac{\pi^2}{8m_\phi^3}\Gamma_{\phi\rightarrow gg}$ and $\tau=\frac{m_\phi^2}{\mathbf{S}}$ with $\mathbf{S}$ be the hadronic centre of mass energy. Here $z_1$ and $z_2$ are the longitudinal momentum fractions carried by the partons 
and $\mathcal{G}(z)$ is the gluon PDF. The luminosity function can be written as, 
 \begin{align}
     \frac{d\mathcal{L}^{gg}}{d\tau}  = \int_{\tau}^{1} \frac{dz}{z} \mathcal{G}(z) \mathcal{G}(\tau/z)~.
 \end{align}
Clearly, both
$\hat{\sigma}(gg\rightarrow \phi)$ and the hadronic $\sigma_{gg\phi}$
depend 
on the $\phi hh$ vertex $\Lambda$. In the above, $\sigma_{\phi}^{\rm LO}$,
$\sigma_{gg\phi}$, and the
luminosity function increase with decreasing mass for the singlet-like state
which helps for the enhancement of $\sigma_{\phi}^{\rm NLO}$. 
A large $\Lambda$ would enhance the two-loop contribution.
Since $\Lambda$ can be constrained from the LHC results, it is instructive to study specific models to make an estimate of the two-loop contribution to the total cross-section.

\section{PRODUCTION OF A LIGHT SCALAR in the extensions of the SM }  
 \label{sec:model}
We now discuss two simple but popular models---(i)
a real scalar extension to the SM ($\varphi$+SM) and (ii) the NMSSM. 
\vskip 0.25cm
(i) {\bf $\varphi$+SM scenario}~: 
In a real singlet extension of the SM, the most general scalar potential which accommodates all the renormalizable terms involving $\varphi$ and the SM Higgs doublet is given by,
\begin{equation}
  V(H,\varphi) = - m_H^2 H^\dagger H - \overline{m}_\varphi^2 \varphi^2 + \lambda (H^\dagger H)^2 + {a_1} H^\dagger H \varphi + {a_2} H^\dagger H \varphi^2 +
\kappa_1 \varphi + {\kappa_3} \varphi^3 + {\kappa_4} \varphi^4~.
\label{scpotential}
\end{equation}
For this tree-level potential, the stability of the vacuum can be ensured if the potential does not become negative along any direction in the field space, which implies
$(i)~\lambda>0, (ii) ~\kappa_4 >0 ~\rm and ~(iii)~
a_2 + 2\sqrt{\lambda \kappa_4} >0, $
along the  $\varphi = 0,~H = 0, $ and $\sqrt{\lambda}H^\dagger H = \sqrt{\kappa_4}\varphi^2 $ directions, respectively. The neutral component of $H$ is denoted as $H_0 = (h^\prime+v)/\sqrt{2}$ with the vacuum expectation value (vev)  being $\langle H_0 \rangle = \frac{v}{\sqrt{2}}$. 
We similarly write $\varphi=\phi^\prime+x$, where the vev of $\varphi$ is denoted as $x$. All the couplings are real here. 

After EWSB, the potential in Eq.~\eqref{scpotential} can lead to two non-trivial scenarios in terms of the vevs
---(i)~$x\neq 0$ and (ii)~$x=0$ with 
$v=246$~GeV.
Our main observations, as will be discussed below, would not have any direct dependence on
whether we assume ~$x\neq 0$ or ~$x=0$. The mass matrix in 
the $(\phi^\prime\ \ h^\prime)^T$ basis can be written as,
\begin{align}
\mathcal{M}_{\phi h}=\left(\begin{array}{c c}
-\overline{m}^2_\varphi+\frac{a_2}{2}v^2 +3 \kappa_3 x +6 \kappa_4 x^2 & a_1 v + 2 a_2 v x\\
a_1 v + 2 a_2 v x& \frac{1}{2}(-m^2_H+3 \lambda v^2 + a_2x^2+a_1x)
\end{array}\right)~.
\label{eq:Mph}
\end{align}
We can rotate $(\phi^\prime\ \ h^\prime)^T$ to a physical basis $(\phi\ \ h)^T$ such that, 
\begin{align}
\left(\begin{array}{c}
\phi^\prime\\
h^\prime
\end{array}\right)=\left(\begin{array}{c c}
\cos\theta & -\sin\theta\\
\sin\theta & \cos\theta
\end{array}\right)\left(\begin{array}{c}
\phi\\
h
\end{array}\right)
\label{eq:U}
\end{align}
In this new basis, the diagonal mass matrix is defined as $\mathcal{M}_D=U^\dagger \mathcal{M}_{\phi h}U$, where $U$ is defined as $(\phi^\prime\quad h^\prime)^T = U (\phi\quad h)^T$. The physical masses are given by,
\begin{align}
m^2_{\phi,h}=
\left(\pm(\overline{M}_\phi^2-\overline{m}^2)(1+2 \tan^2\theta)+(\overline{m}^2+ \overline{M}^2_\phi )\right)~.
\label{Eq:eigen}
\end{align}
Here $m_h=125~$GeV; $\overline{M}^2_\phi,\overline{m}^2$ are the diagonal elements of $\mathcal{M}_{\phi h}$ with the mixing angle,
\begin{align}
\tan\theta=a_1 v + 2 a_2 v x/\left(\overline{M}_\phi^2-\overline{m}^2\right).
\label{Eq:mix}
\end{align}

For $\phi$ to be mostly singlet and $h$ to be SM-like, the mixing angle $\theta$ has to be small.
In this work, we are interested in the parameter space where $\phi$ is the lightest physical scalar with $m_\phi < m_h$ and $|\sin\theta| \le 0.2$ \cite{ATLAS:2020qdt}. 
Following Eq.~\eqref{scpotential}, we observe that the $\phi hh$ coupling is proportional to $(a_1 + 2 a_2 x)$. Since the same factor 
also appears in $\tan\theta$, we may redefine
the coupling as $a_1 \equiv (a_1 + 2 a_2 x)$. Thus, for simplicity, we   
take $a_2 \to 0$ in Eq.~\eqref{scpotential}. This 
helps us to avoid the constraints 
on the $h\to \phi\phi$ decay for a light $\phi$.
Then the effective $\phi hh$ interaction simply becomes 
$ \simeq - \frac{a_1}{2} \phi hh$ (with $\cos\theta \to 1$). Comparing with 
the generic $\phi hh$ vertex
$\Lambda$, this yields $|\Lambda| = \frac{a_1}{2}$. 

\begin{figure}[!h]
	\centering
	\includegraphics[width=0.485\linewidth]{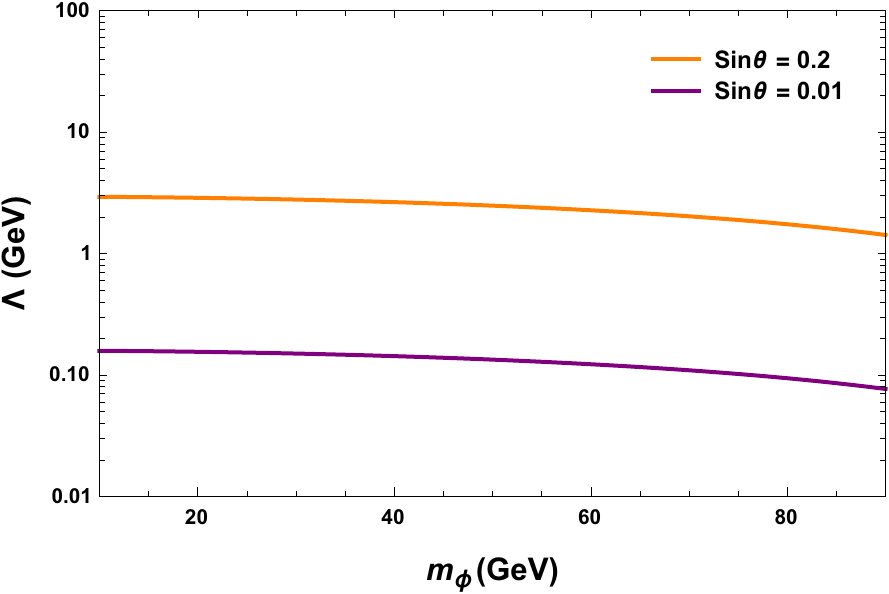}
		\includegraphics[width=0.50\linewidth]{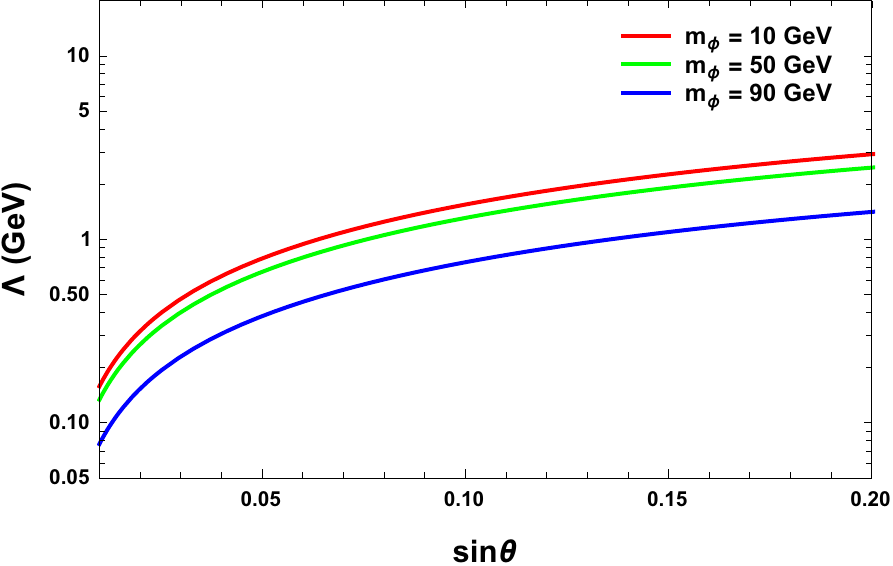}
		\hskip 4cm(a)\hskip 8cm (b)
	\caption{Variations of $\Lambda$ with $m_{\phi}$ and $\sin\theta$ in the $\varphi$+SM.}
\label{fig:plotlambda2}
\end{figure}

Now, Eqs.~\eqref{Eq:eigen} and~\eqref{Eq:mix} can be solved for given values of $m_\phi$ and $\tan\theta$. As a result, the variation in $\Lambda$ with $m_{\phi}$ or $\sin\theta$ can  be easily obtained (see, e.g., Fig.~\ref{fig:plotlambda2}). We focus on the scenario where $\theta$ is small so that $\tan\theta \simeq \sin\theta$. In Fig.~\ref{fig:plotlambda2}a, we show the variations in $\Lambda$ with $m_{\phi}$ for $\sin\theta = 0.2,~0.01$, while in Fig.~\ref{fig:plotlambda2}b, the variation is 
with $\sin\theta$. For the latter, three representative values of $m_\phi$ are considered. We see that
here $\phi hh$ can only take a much smaller value, $\Lambda \sim 3$. Moreover, for a small $\sin\theta$, $\Lambda$ also becomes small, reducing the two-loop contributions for $\phi$ production.

A more general scenario can have a complex singlet $\Phi$ instead of the real singlet, which can lead to the production of a CP-odd scalar in addition to its CP-even counterpart. Let $\mathcal{A}$ refers to the CP-odd component of $\Phi$ in the physical basis. The amplitude for $ gg \to \mathcal{A}$
can be obtained if one replaces one of the $h$ in Fig. \ref{fig:ggphi_two_loop} with $\mathcal{A}$.
However, for a dominantly singlet-like state, $\mathcal{A} t \bar t$ is much smaller compared to the $h t \bar t$ coupling. An enlarged Higgs sector, e.g., the addition of $\Phi$ to two-Higgs doublet  
 models can be helpful, as a doublet-like CP-odd state can boost the couplings to the SM fermions. This will in turn lead to a larger amplitude from the $\mathcal{A} gg$ effective coupling. Similarly, for the CP-even states, it may be possible to obtain a larger value of $\Lambda$ satisfying
the LHC constraints, since additional parameters are involved in the mixing matrix. \\
\vskip 0.1cm

(ii) \textbf {NMSSM and a large value of $\Lambda$}~: 
NMSSM as a part of an extended two-Higgs doublet model consists of three complex scalars $H_u^0$, $H_d^0$, and $S$, where $S$ is the gauge singlet \cite{Ellwanger:2009dp}. In the $Z_3$ invariant NMSSM, their self-couplings can be traced from the superpotential $W$ in terms of the superfields ${\hat H_{u,d},~\hat S}$, and the trilinear soft supersymmetry breaking terms.
\begin{equation}
W=\lambda {\hat H}_u { \hat H}_d { \hat S} + \frac{\kappa}{3} {\hat S}^3 + \dots \quad ; \quad L_{soft} \supset(\lambda A_\lambda  H_u  H_d  S + \frac{\kappa}{3} A_\kappa S^3) + \text{h. c.}\; .
\end{equation}

\begin{figure}[!h]
	\centering
	\includegraphics[width=0.49\linewidth]{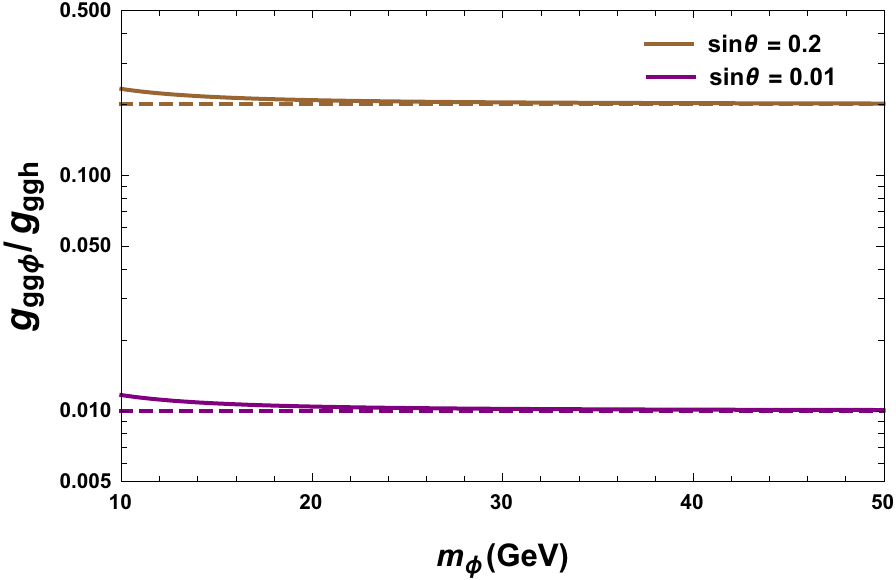}
		\includegraphics[width=0.487\linewidth]{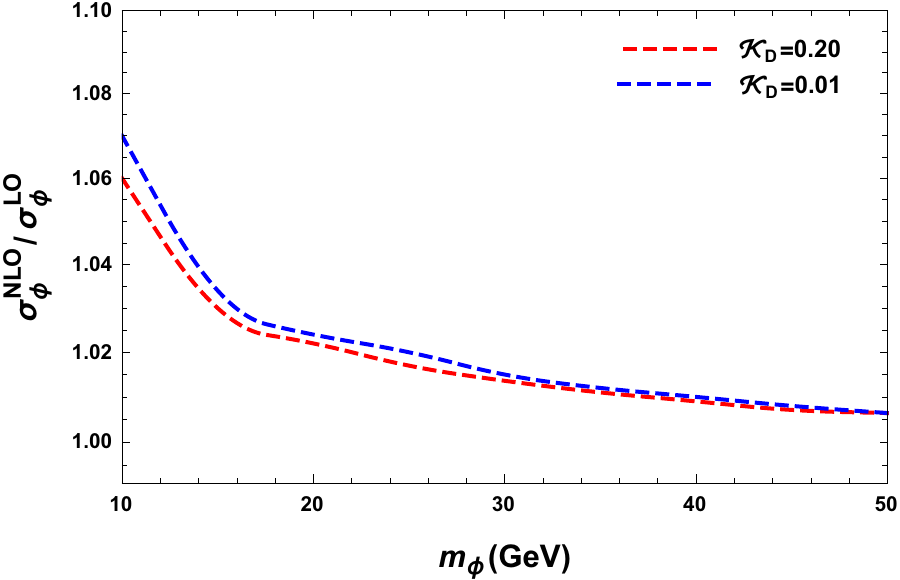}\\
		\hskip 1.0cm(a)\hskip 7.2cm (b)
		\caption{(a) We present contours for the reduced couplings 
	that include the total contributions 
	(${g_{ gg \phi}}/{g_{gg h}}$) (solid lines) and 
	${g^D_{ gg \phi}}/{g_{ggh}}$ (dashed lines)
	with $m_{\phi}$. 
	Vertical shifts reflect the contributions coming from two-loop parts which become important for $m_\phi \le 30$ GeV. (b) Variations of $\frac{\sigma_\phi^{\rm NLO}}{\sigma_\phi^{\rm LO}}$ with $m_\phi$ in $\phi+$SM. For smaller values of
                $\mathcal{K}_D$, singlet part in $\phi$ increases, thus NLO corrections seem to enhance.}
        \label{fig:plot2}
\end{figure}

Due to the presence of additional fields and couplings in the Higgs sector, it is possible to have a larger value for the $\phi h h$ coupling while keeping the $\phi$ couplings to the SM fermions small. This lowers the one-loop $g g \phi $ amplitude mediated by $g^D_{ gg \phi}$ (see Eq.~\eqref{Eq:effcoup}).
Note that, it is difficult to obtain in the simple $\varphi+$SM scenario. Here, in the Higgs basis, assuming, $s,A_\lambda \gg v_{u,d}(\sim M_Z)$ ($s$ being the vev of $S$), $ \phi h h$ coupling can be read as,
\begin{align}
\Lambda \sim \frac{\lambda}{\sqrt{2}} \mu - \frac{\lambda}{\sqrt{2}}\sin\beta \cos\beta (2 \frac{\kappa \mu}{\lambda} + A_\lambda)\;, 
\label{Eq:nmssm_phihh}
\end{align}
with $\mu = \lambda s$. Numerically, $\lambda,\ \kappa,\,\,\,\, A_\lambda,\ A_\kappa,\ \mu,\ \tan\beta\equiv {v_u}/{v_d}$ can be chosen to frame a few benchmark points (BMPs) allowed by the $\Lambda$ and LHC constraints. This article uses {\sc NMSSMTools }\cite{Ellwanger:2004xm,Ellwanger:2005dv,Das:2011dg} to make sure all the relevant LHC constraints are respected by our numerical analyses. 
\vskip 0.2cm
Finally, we present the results for the two-loop hadronic cross-sections $\sigma_\phi^{\rm NLO}$ at the 14 TeV LHC using the NNPDF30 LO AS 0118 \cite{NNPDF:2014otw} PDF. We use Eqs.~\eqref{ggphi_cross_sec} and~\eqref{eq:had} for the calculation. First, we look for the regions in the parameter space of $m_\phi$ where the two-loop NLO amplitude $g^S_{ gg \phi}$ can lead to some enhancement to the LO 
amplitude. The results can be seen from Fig.~\ref{fig:plot2} where we plot the contours for ${g_{gg \phi }}/{g_{gg h}}$ (solid lines) with $m_\phi$. Notably, $g_{gg \phi}$ includes both LO and the NLO amplitudes where the corresponding LO values are given by ${g^D_{gg \phi }}/{g_{gg h}}$. The ratio denoted as the reduced coupling ${\mathcal K}_D$ (see Eq.~\eqref{Eq:effcoup}) or $\sin\theta$,
has been fixed at 0.2 and 0.01 for illustration. For these values of $\sin\theta$, the only relevant quantity for the two-loop amplitude, $\Lambda$, can be calculated for the $\varphi$+SM from Eqs.~\eqref{Eq:eigen} and~\eqref{Eq:mix}. Though there can be a mild dependence of $\Lambda$ over $m_\phi$, we ignore the effect and set $\Lambda=3, 0.15$ GeV for $\sin\theta=0.2, 0.01$ respectively. Clearly, the NLO amplitude would cause a vertical shift in ${g_{gg \phi }}/{g_{gg h}}$ at small values of $m_{\phi}$ compared to the flat LO value set by $\sin\theta$ (see Fig.~\ref{fig:plot2}).
 This is also reflected in Fig.~\ref{fig:plot2} where $\frac{\sigma_\phi^{\rm NLO}}{\sigma_\phi^{\rm LO}}$ with $m_\phi$
  has been presented for $\mathcal{K}_D$=0.2 and 0.01. Note that when $\phi$ becomes more singlet-like, say $\mathcal{K}_D=0.01$, the boost in the total cross-section increases as compared to the value evaluated for $\mathcal{K}_D=0.2$. The NLO corrections become less important for $m_\phi>$30 GeV. Though this is particularly true in the $\varphi$+SM scenario, the situation might improve in an extended Higgs framework like the NMSSM. We will see that a relatively heavy singlet can still receive a tiny but nonzero contribution from the NLO diagrams.

We present the NLO-corrected cross-sections $\sigma^{\rm NLO}_\phi$ for a few benchmark points (BMPs) for the two models $\phi+$SM and the NMSSM, in Table~\ref{tab:my_label}. The LO cross-section has been borrowed from the NNLO+NNLL QCD prediction at the 14 TeV LHC for a Higgs state having SM-like couplings, 
$\sigma_{ggh}^{\rm SM}\simeq 49.47$ pb \cite{Higgs14}. The cross-section is subsequently scaled by ${\mathcal K}^2_D$ and tabulated as $\sigma_{\phi}^{\rm LO}$. Other NLO EW correction factors are not 
included except those mediated by the $\phi hh$ vertex, $\sigma_{gg\phi}$, and its interference with the LO term. In the $\varphi+$SM, non-negligible contributions can be obtained only for a light $\phi$, which
amounts to a $\sim 7\%$ enhancement to the one-loop value. For $m_\phi \geq 30$ GeV, the NLO contributions become insignificant to LO results. However, the situation improves with the presence of additional Higgs in the spectrum. For example, in the NMSSM, we see a minimal boost of $\sim 2\%$ for a relatively heavier $\phi$. Here, a relatively large $\Lambda$ can be obtained while keeping the $g^D_{gg \phi }$ at a very small value, which in turn may cause a reduction in $\phi$ production via one-loop. In the NMSSM, we consider $m_\phi > \frac{m_h}{2}$ to prevent the $h\to\phi\phi$ decay.
\begin{table}[h!]
	\centering
	\begin{tabular}{ |p{1.4cm}|p{1.2cm}|p{1.3cm}|p{1.2cm}|p{1.7cm}|p{1.6cm}|p{1.3cm}|  }
		\hline
		\multicolumn{7}{|c|}{Cross section for $pp\rightarrow \phi$ at $\sqrt{\mathbf{S}}=14$ TeV} \\
		\hline
		\,\,\,Model &\,\,\,\,\,\,$\mathcal{K}^2_D$&$m_\phi$(GeV) & $\Lambda$(GeV) & \,\,\,$\sigma_{\phi}^{\rm LO}$ (pb)& \,\,$\sigma_{\phi}^{\rm NLO}$(pb) &\,\, $\displaystyle\frac{\sigma_\phi^{\rm NLO}}{\sigma_\phi^{\rm LO}}$\\
		\hline
		\hline
		
		$\phi$+SM &0.0400  &\,\quad10 &\,\,\,\,2.92  &\, 297.6400      &\, 315.4320  &\,\,\,\,\,1.060    \\
		\cline{3-7}
		& & \,\quad15 &\,\,\,\,2.90 &\,183.0400 &\, 188.6070\,&\,\,\,\,\,1.030\\
		\cline{3-7}
		& & \,\quad20 &\,\,\,\,2.87  &\, 89.8000 &\, 91.7988 &\,\,\,\,\,1.022  \\
		\cline{3-7}
		& & \,\quad25 &\,\,\,\,2.82  &\, 49.6800 &\, 50.5200 &\,\,\,\,\,1.017  \\
		\cline{2-7}
		&  0.0001  & \,\quad10 &\,\,\,\,0.16    &\, 0.7441      &\, 0.7934 &\,\,\,\,\,1.070   \\
		\cline{3-7}
		&  & \,\quad15 &\,\,\,\,0.16  &\, 0.4576        &\,0.4732 &\,\,\,\,\,1.034 \\
		\cline{3-7}
		&  & \,\quad20 &\,\,\,\,0.15  &\, 0.2245       &\, 0.2300 &\,\,\,\,\,1.024 \\
		\cline{3-7}
		&  & \,\quad25 &\,\,\,\,0.15  &\, 0.1242       &\, 0.1266 &\,\,\,\,\,1.020 \\
		\hline
		\hline
		NMSSM  &0.0079 &\,\quad70 &\,\,\,\,6.00  &\, 1.0278        &\, 1.0462 &\,\,\,\,\,1.020  \\
		\cline{2-7}
		&0.0046 &\,\quad80 &\,\,\,\,5.93 &\, 0.4730       &\, 0.4820 &\,\,\,\,\,1.019  \\
		\cline{2-7}
		&0.0016 &\,\quad90 &\,\,\,\,4.72 &\, 0.1340       &\, 0.1370 &\,\,\,\,\,1.022  \\
		\cline{2-7}
		\hline
	\end{tabular}
    \caption{Production cross-section of $\phi$ in the $\varphi+$SM and the NMSSM. For $\varphi$+SM, 
    a light scalar is preferred while for a relatively heavy scalar NMSSM shows a non-negligible effect.}
    \label{tab:my_label}
    \end{table}
    \section{Conclusions}
\label{conclusion}
We have calculated the two-loop EW $gg\phi$ amplitude mediated by the $\phi hh$ vertex for a singlet-like state.
After expressing the two-loop amplitude through one-loop effective vertices $gghh$ and then $gg\phi$, we have
explicitly calculated the total NLO correction from the two diagrams. We have found
the two-loop $gg\phi$ amplitude to be UV-finite and independent of the choice of the renormalization scale.
Considering two simple models (i) $\varphi+$SM and (ii) NMSSM, we have calculated the two-loop NLO cross-sections.
In $\varphi+$SM, the two-loop contribution for a light $\phi$ can lead up to a $\sim 7\%$ increase in the total
cross-section. In the NMSSM, one can see a relatively smaller but nonzero enhancement in the total cross-section
for the lightest CP-even singlet-like scalar. NLO corrections yield better results when $\phi$
  becomes more singlet-like. Our results are general and can easily be
used in other models with enlarged scalar sectors if the coupling $\phi h h$ is known.

\section{Acknowledgements}
Our computations were supported in part by SAMKHYA:
the High-Performance Computing Facility provided by the
Institute of Physics (IoP), Bhubaneswar, India. We acknowledge P. Agrawal, P.P. Giardino, K. Ray, D. Maity, A. Bhaskar, and B. Das for valuable discussions.
\vfill
\appendix
\section{}
\label{appedxixA}
The coefficients $\xi_1^{(1)},...,\xi^{(1)}_{12}$, $\xi_1^{(2)},...,\xi^{(2)}_{12}$ are given by,
\begin{align}
   \xi^{(1)}_1 =&  \frac{2}{m_\phi^2}\Big[12m_t^2m_\phi^2 + (2-d)m_\phi^4 + 2(d-4-2x+xd)m_\phi^2(k.\ell)+ 8x(1+x)(k.\ell)^2 + 2m_\phi^2x(d-2xd-3\nonumber\\
   &+6x)(\ell.\ell) + 2\bigl\{4m_\phi^2-dm_\phi^2-2m_\phi^2x+dm_\phi^2x-4x(k.\ell)-4x^2(k.\ell)\bigr\}(\ell.q)\Big]~,\\
   \xi^{(1)}_2 =& -4(d-3)~,\\
   \xi^{(1)}_3 =& \frac{2}{m_\phi^2}\big[-8(k.\ell)+ 2\bigl\{4x(2-x)(\ell.\ell)-4(3m_t^2+\ell.q)\bigr\}\Big]~,\\
   \xi^{(1)}_{4} =& \frac{2}{m_\phi^2}\Big[24m_t^2 +8x(x-2)(\ell.\ell)+8(\ell.q)\Big]~,\\
   \xi^{(1)}_{5} =& \frac{16(k.\ell)}{m_\phi^2}~,\\
   \xi^{(1)}_6 =& -\frac{16}{m_\phi^2}~,\\
   \xi^{(1)}_7 =& \frac{16}{m_\phi^2}~,\\
   \xi^{(1)}_8 =& 0~,\\
   \xi^{(1)}_9 =& \frac{2}{m_\phi^2}\Big[4m_\phi^2(d-3)+8x(1-2x)(k.\ell)\Big]~,\\
   \xi^{(1)}_{10} =&  8x(d-3+6x-2xd)~,\\
   \xi^{(1)}_{11} =& \frac{2}{m_\phi^2}\Big[16x(2x-1)(k.\ell)+4\bigl\{m_\phi^2(3-d-2x+xd)+2x(1-2x)(\ell.q)\bigr\}\Big]~,\\
   \xi^{(1)}_{12} =& \frac{1}{m_\phi^2}\Big[16x\bigl\{m_t^2(3x-1)+x^2(x-1)(\ell.\ell)\bigr\}(k.\ell)^2 + 4\bigl\{m_t^2m_\phi^2(4-d-10x+3xd)+4xm_t^2(1-3x)(\ell.q)\nonumber\\
   &+x^2(x-1)(dm_\phi^2-2m_\phi^2-4x\,\ell.q)(\ell.\ell)\bigr\}(k.\ell)-2m_\phi^2\bigl\{2(1-d)m_t^4 + (d-2)m_t^2m_\phi^2+ 2x^3(x-1)(d\nonumber\\
   &-3)(\ell.\ell)^2+2m_t^2(x-1)(d-2)(\ell.q)+ 2m_t^2x(1+d-6x)(\ell.\ell)+x(x-1)m_\phi^2(d-2)(\ell.\ell)-2x^2(x\nonumber\\
   &-1)(d-2)(\ell.q)(\ell.\ell)\bigr\}\Big]~,\\
    \xi^{(2)}_1 =& \frac{2}{m_\phi^2}\Big[4m_t^2m_\phi^2(5-d)+(2-d)m_\phi^4+2m_\phi^2(d-2+4x-2xd)(k.\ell)-8(k.\ell)^2+2m_\phi^2(d-5+2x-2xd\nonumber\\
    &-2x^2+2x^2d)(\ell.\ell)+ 2m_\phi^2(4-d-2x+2xd)(\ell.q)+ 8(k.\ell)(\ell.q)\Big]~,\\
    \xi^{(2)}_2 =& 4(d-1)~,\\
    \xi^{(2)}_3 =& \frac{2}{m_\phi^2}\Big[-16(k.\ell)+ 8(\ell.\ell)-8\bigl\{4m_t^2+(\ell.q)\bigr\}\Big]~,\\
    \xi^{(2)}_4 =& \frac{2}{m_\phi^2}\Big[32m_t^2-8(\ell.\ell)+16(\ell.q)\Big]~,\\
    \xi^{(2)}_5 =& \frac{16(k.\ell)}{m_\phi^2}~,
    \end{align}\clearpage
    \begin{align}
    \xi^{(2)}_{6,7,8} =& 0~,\\
    \xi^{(2)}_9 =& \frac{2}{m_\phi^2}\Big[16(x-1)(k.\ell)-4m_\phi^2(1+2x-xd)\Big]~,\\
    \xi^{(2)}_{10} =& 16\Big[1-x(d-1)+x^2(d-1)\Big]~,\\
    \xi^{(2)}_{11} =& \frac{2}{m_\phi^2}\Big[16(k.\ell)+4\bigl\{m_\phi^2(d-2+4x-2xd)-4x(\ell.q)\bigr\}\Big]~,\\
    \xi^{(2)}_{12} =& -\frac{2}{m_\phi^2}\Big[(d-2)m_t^2m_\phi^4 -2(d-1)m_t^2m_\phi^2-8m_t^2(1-2x)^2(k.\ell)^2 + 2\bigl\{m_t^2m_\phi^2(d-2+4x-2xd)+ 4m_t^2\nonumber\\
    &\times(1-2x)^2(\ell.q)+x(x-1)m_\phi^2(d-2)(2x-1)\bigr\}+m_\phi^2\bigl\{-2x^2(x-1)^2(d-1)(\ell.\ell)^2+2m_t^2(x-1)\nonumber\\
    &\times(d-2)(\ell.q)+m_\phi^2x(x-1)(d-2)(\ell.\ell)+2m_t^2(d-3+10x-10x^2-2xd+2x^2d)(\ell.\ell)-2x^2(x\nonumber\\
    &-1)(d-2)(\ell.q)(\ell.\ell)\bigr\}\Big]~.
\end{align}

\bibliographystyle{JHEPCust.bst}
\bibliography{phi2loop}
\vfill
\end{document}